\RequirePackage{luatex85} 
\documentclass[aip, amsmath,amssymb, reprint]{revtex4-1}

\usepackage{amssymb,amsfonts,amsmath}
\usepackage{hyperref} 
\usepackage{siunitx}
\usepackage[normalem]{ulem}
\usepackage[colorinlistoftodos]{todonotes}
\usepackage{stmaryrd}

\definecolor{brown}{rgb}{0.59, 0.29, 0.0}
\definecolor{orange}{RGB}{255,127,0}
\definecolor{brightube}{rgb}{0.82, 0.62, 0.91}

\graphicspath{{Figs/}}

\definecolor{darkgreen}{rgb}{0, 0.5, 0.05}

\newcommand{\deleted}[1]{}

\newcommand{\eqn}[1]{\begin{align}#1\end{align}}
\newcommand{\bs}[1]{\boldsymbol{#1}}

\newcommand{\pare}[1]{\left( #1 \right) }

\newcommand{\fr}[2]{\frac{#1}{#2}}

\newcommand{\mc}[1]{\mathcal{#1}}
\newcommand{\avg}[1]{\langle #1 \rangle}
\newcommand{\tex}[1]{\mbox{\scriptsize{#1}}}

\newcommand{\what}[1]{\widehat{#1}}

\def\dt{\Delta t}
\def\dd{\mathrm{d}}  
\def\kt{k_B T}
\def\bna{\bs{\nabla}}
\def\dash{\text{-}}

\def\bbf{\bs{f}}
\def\bF{\bs{F}}
\def\bG{\bs{G}}

\def\bK{\bs{K}}

\def\bM{\bs{M}}

\def\bq{\bs{q}}

\def\br{\bs{r}}

\def\bu{\bs{u}}
\def\bU{\bs{U}}
\def\bv{\bs{v}}

\def\bW{\bs{W}}
\def\bx{\bs{x}}

\def\by{\bs{y}}

\def\btau{\bs{\tau}}
\def\bomega{\bs{\omega}}
\def\blambda{\bs{\lambda}}

\def\bmW{\bs{\mc{W}}}

\def\bzero{\bs{0}}
\def\gammaDot{\dot{\gamma}}

\begin{document}

\title{Rheology of moderated dilute suspensions of star colloids: the shape factor}


\author{F. Balboa Usabiaga}
\affiliation{BCAM - Basque Center for Applied Mathematics, Alameda de Mazarredo 14, E48009 Bilbao, Basque Country - Spain}

\author{M. Ellero}
\affiliation{BCAM - Basque Center for Applied Mathematics, Alameda de Mazarredo 14, E48009 Bilbao, Basque Country - Spain}
\affiliation{Ikerbasque, Basque Foundation for Science, Calle de Maria Diaz de Haro 3, E48013 Bilbao, Basque Country - Spain}
\affiliation{Zienkiewicz Center for Computational Engineering (ZCCE), Swansea University, Bay Campus, Swansea SA1 8EN, UK}

\begin{abstract}
  Star colloids are rigid particles with long and slender arms connected to a central core.
  We show numerically that the colloid shapes control the rheology of their suspensions.
  In particular, colloids with curved arms and \emph{hooks} can entangle with neighbor particles and form large clusters that can sustain high stresses.
  When a large cluster permeates the whole system the viscosity increases many fold.
  Contrary to the case of spherical colloids we observe that these effects are very strong even at moderate volumes fraction over a wide range of Péclet numbers.  
\end{abstract}
\maketitle

\maketitle


\section{Introduction}
\label{sec:introduction}

The rheology of colloidal suspensions has been studied extensively for spherical particles \cite{Kruif1985, Doi1994, Segre1995, Foss2000, Sierou2001, Cwalina2014}.
The case of non-spherical colloidal suspensions has also been studied for relatively simple shapes such as ellipsoids and fibers \cite{Mewis2012}
or spheres with roughness \cite{Moon2015, Tanner2016}.
The experimental work date back, at least, to the 1940s when Lauffer employed tobacco viruses to study the rheology of suspension of rod-like colloids.
For rod-like colloids the viscosity increases with the colloidal aspect ratio and for very dilute suspensions
the viscosity can be significantly larger than for suspension of spherical colloids at the same volume fraction \cite{Lauffer1944}.
The stress generated in dilute suspensions of axisymmetric particles such as spheroids and fibers was studied by Brenner who gave explicit analytical expressions \cite{Brenner1974}.
Such axisymmetric colloids generate stresses that depend on their orientation which in turn introduce viscoelasticity effects even in suspensions with a Newtonian matrix
and show a strong shear thinning \cite{Brenner1974, Nemoto1975}.
The rheology of non-spherical colloids has also been studied computationally for colloids with relatively simple shapes such as platelets \cite{Meng2008, Meng2008a, Yamamoto2012}, spheroidal particles \cite{Bertevas2010, Trulsson2018} or dimeric colloids \cite{Mari2020}.

However, many applications and natural systems contain colloids with more complex shapes.
For example, marine snow is formed by a complex aggregate of organic particles
which can range from globular to rod-like particles and even long filaments and sheets \cite{Logan1990, Alldredge1988}.
The interactions and entanglement between such particles can affect the aggregates rheology, cohesion and dynamics \cite{Guadayol2021, Song2022, Trudnowska2021}.
Another example is aluminum alloy melts where iron impurities form complex-shaped colloids that intertwine into clusters
that contaminate the alloys and can worse their mechanical properties \cite{Caceres2006}.
Additionally, complex-shaped colloids can now be carefully synthesized thanks to the new developments in colloidal science.
For example, $\text{SiO}_2$ can be grown over a colloidal hematite to design colloids with diverse shapes such as cubes or peanut-like shells \cite{Sacanna2011a}.
Controlled polymerization over a seed and further treatment of the colloids can produce diverse shapes such as dumbbells, barrels or spheres with concavities \cite{Sacanna2011a}.
Photolithography has also been used to synthesize plane colloids of diverse shapes with high accuracy \cite{Chakrabarty2013a}.
Further treatment of such colloids can produce even more complex shapes such as Janus particles with tree-like shapes with thin and elongated branches \cite{Dai2016}.
Another fabrication approach is the controlled aggregation of smaller colloids to form, for example, plane star shape colloids \cite{Lapointe2013} or amorphous aggregates with slender arms \cite{Bourrianne2022}.

Complex-shaped colloids have distinct interactions that can affect their rheology.
For example, cubic colloids show a strong shear thickening at steady and oscillatory flows
as their lubrication forces are stronger than the ones between spherical colloids \cite{Cwalina2016, Cwalina2016a}.
Bourrianne et al. found that suspensions of amorphous colloids formed by the aggregation of small spherical silica particles
yield continuous and discontinuous shear thickening (CST and DST) at volume fractions as low as $\phi \sim 10\%$ \cite{Bourrianne2022}.
Bourrianne et al. showed that the DST was controlled by the levels of hydrogen bonding interactions between colloids \cite{Bourrianne2022}.
A result that can be cast into the frictional model introduced by Seto et al. to explain DST in suspensions of spherical colloids \cite{Seto2013, Mari2015}.

Here, we study the rheology of a suspension of star colloids, also called Czech hedgehog colloids, which are formed by slender arms connected to a central core \cite{Westwood2022}.
They can be seen as an idealization of the amorphous colloids used in some experiments like the ones of Bourrianne et al \cite{Bourrianne2022}. 
However, we do not include any friction model to incorporate the hydrogen bonds between colloids.
Instead, we focus on the role of the colloidal morphology in the suspension viscosity.
We find that the viscosity increases dramatically with the volume fraction and that the suspensions show shear thinning at moderate Péclet numbers (Pe).
Moreover, we observe that the viscosity depends strongly on the colloidal shape, thus,
the shape could be seen as a new knob to control the rheology of colloidal suspension.

The rest of the paper is organized as follows.
In Sec. \ref{sec:setup} we describe the model used to simulate colloidal suspensions and the morphology of star colloids.
In Sec. \ref{sec:results} we describe our protocol to measure the suspension viscosity and our results.
We show that the suspensions of star colloids show large viscosities at moderate volume fractions and that the shape of the colloids control the viscosity growth with volume fraction.
In particular we show that colloids with curved arms or \emph{hooks} can entangle with nearby colloids to generate a network that resist stresses.
We also study the viscosity dependence with the shear rate and observe shear thinning at moderate Pe numbers.
We finish with conclusions in Sec. \ref{sec:conclusions} and some numerical details in the \ref{sec:spheres}.

\section{Model description}
\label{sec:setup}

\begin{figure}
  \begin{center}
    \includegraphics[width=0.8 \columnwidth]{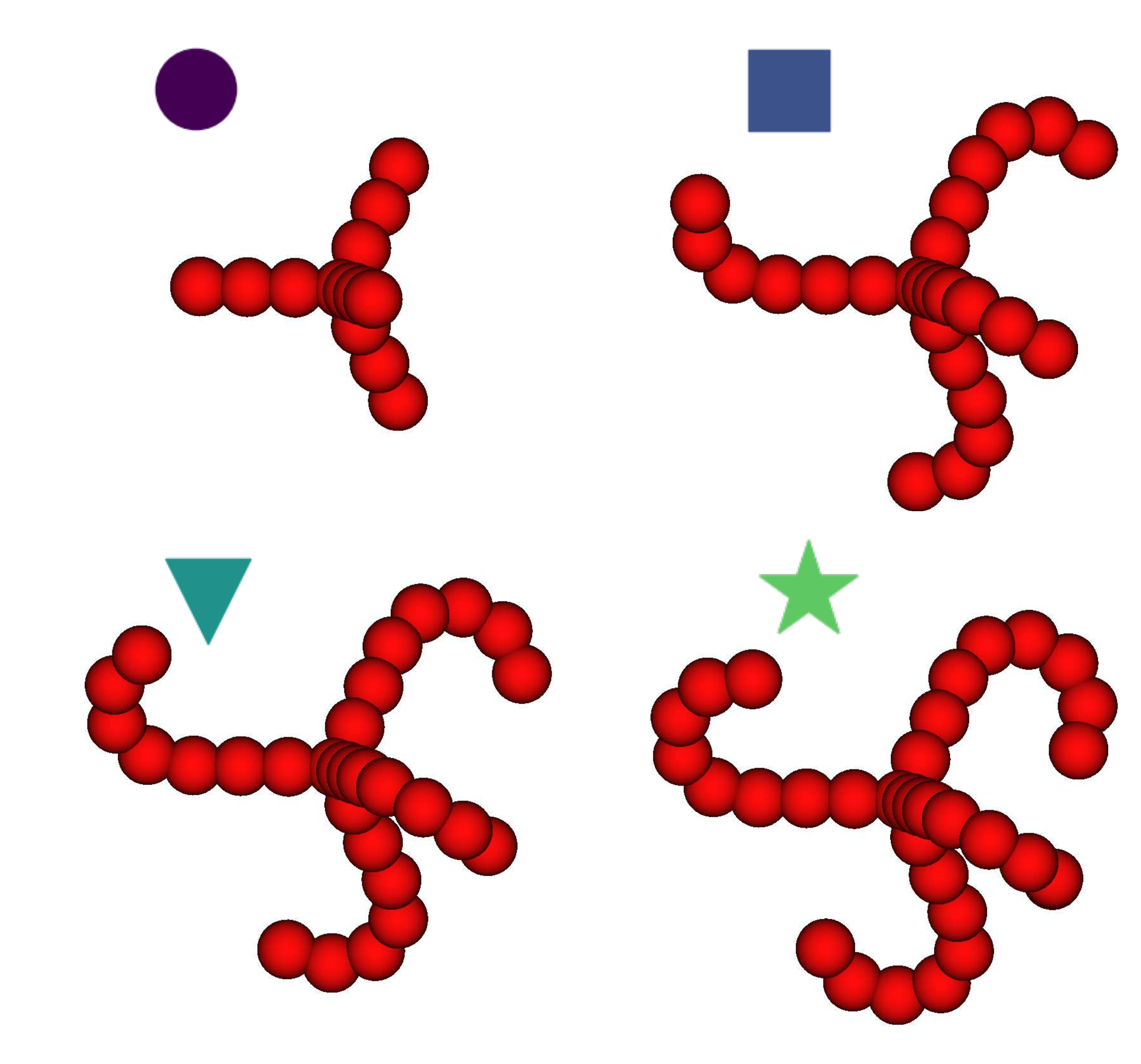}
    \includegraphics[width=0.99 \columnwidth]{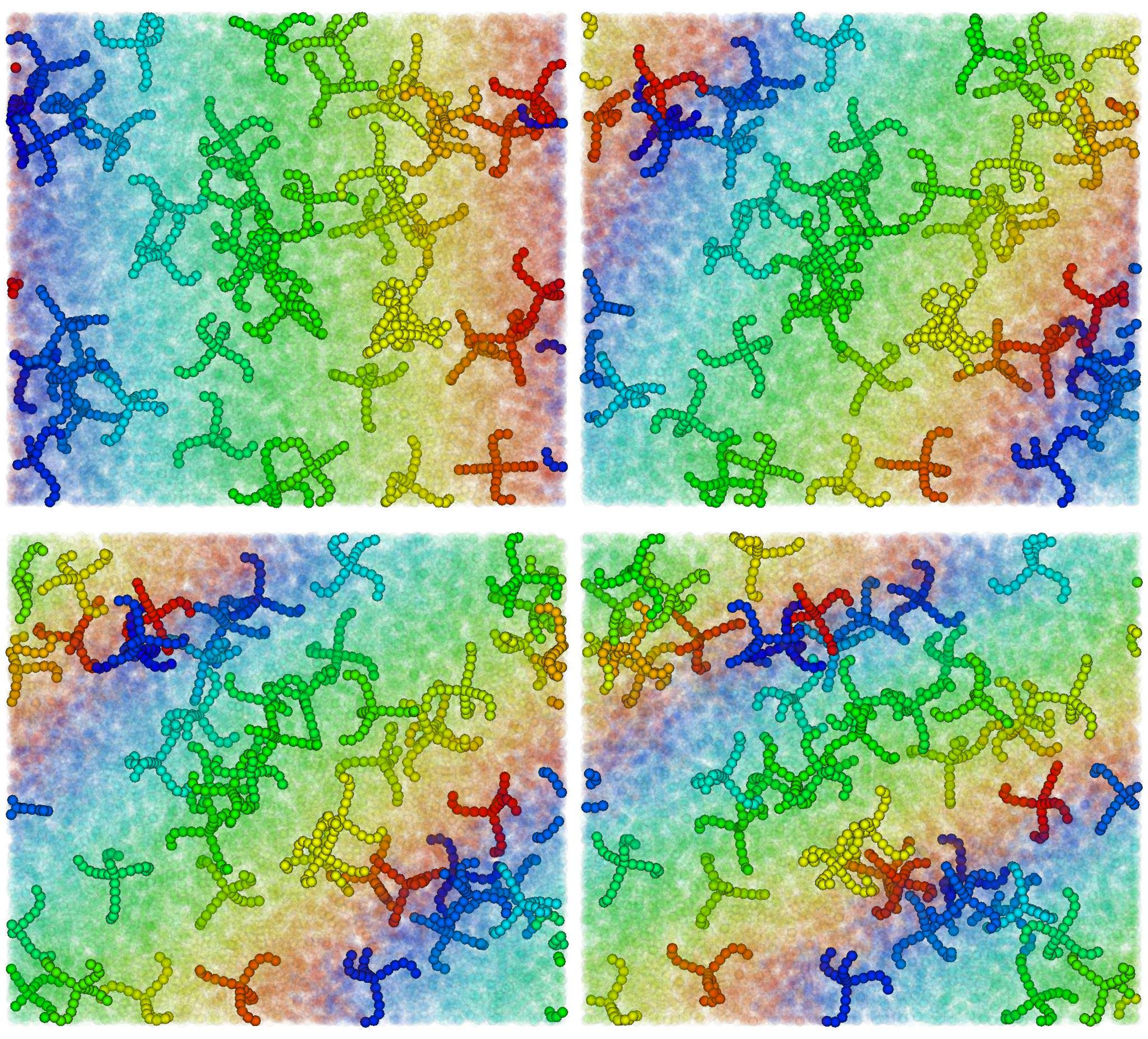}
    \caption{{\bf (Top)} Star colloid models used in this study and the symbols used to label their results in the rest of the paper.
      The colloids have four arms ending with hooks of different lengths.
      Their diameters (maximum distance between arms) go from $d=0.39\si{\mu m}$ for the first model to $d=0.71\si{\mu m}$ for the fourth model.
      {\bf (Bottom)} Snapshots of a suspension under a linear shear flow at volume fraction $\phi=0.2$ and Péclet number $Pe\approx 10$.
      A selected number of colloids is shown with full colors and the rest with a transparency to easy the visualization
      The colloids are colored with their initial position along the x-axis to help visualize the shear motion (Multimedia view). 
    }
    \label{fig:models}
  \end{center}
\end{figure}

We simulate a Brownian suspension with $N$ colloids immersed in a Newtonian fluid. 
The colloids have complex shapes as shown in Fig.\ \ref{fig:models} (Multimedia available online). 

Each colloid has four arms that extend from the center to the vertices of a tetrahedron. 
The arms end with a \emph{hook} shaped as a circular arc. 
In this work we consider four colloid models where only the size of the hooks is varied. 
The hooks subtend an angle $\theta = 0^{\circ} \text{ (no hook)}, 110^{\circ}, 150^{\circ} \text{ and } 180^{\circ}$ 
in each of the four models and the colloidal diameters are $d=0.39,\, 0.69,\, 0.71 \text{ and } 0.71 \,\si{\mu m}$ respectively. 
For simplicity, in this work all the hooks have the same relative orientation as shown in Fig.\ \ref{fig:models}. 
The size of the hooks controls the rheology of the suspension as it will be shown later. 
At small Reynolds numbers the flow is governed by the Stokes equations  \cite{Kim1991, Pozrikidis1992} 
\eqn{
  \label{eq:Stokes}
  -\bna p + \eta_0 \bna^2 \bv &= \sqrt{\fr{2\eta_0\kt}{\dt}} \bna \cdot \bmW^n, \\
  \label{eq:div}
  \bna \cdot \bv &= 0,
}
where $\eta_0$ is the solvent viscosity and $\bv$ and $p$ the flow velocity and pressure.
The term on the right hand side of \eqref{eq:Stokes} introduces the thermal fluctuations acting during the time step $n$ of size $\dt$ \cite{Delong2014a,Bao2018, Sprinkle2019}.
The fluctuations increase with the thermal energy, $\kt$, and are delta correlated in space and time
\eqn{
  \avg{\mc{W}^n_{ij}(\bx) \mc{W}^m_{kl}(\bx')} = \delta_{nm} \pare{\delta_{ik}\delta_{jl} + \delta_{il}\delta_{jk}} \delta(\bx - \bx').
}
At the colloids surfaces the flow obeys the no slip condition \cite{Pozrikidis1992}
\eqn{
  \label{eq:no-slip-continuum} 
  \bv(\br) = \bu_p + \bomega_p \times (\br - \bq_p) \; \mbox{for }\br \in \text{surface colloid } p,
}
where $\bu_p$ and $\bomega_p$ are the linear and angular velocities of colloid $p$ and $\bq_p$ its center.
Additionally, we impose a background linear shear flow,
which is equivalent to the  boundary condition  $\bv(\bx) = \bv_0(\bx) = \gammaDot_0 z \what{\bs{e}}_x$ at infinity,
where $\gammaDot_0$ is the imposed shear rate.
The equations are closed with the balance of force and torque for each colloid $p$ 
\eqn{
  \label{eq:balanceF_continuum}
  \int_{S_p} \blambda(\br) \,\dd S_{\br} = \bbf_p, \\
  \label{eq:balanceT_continuum}
  \int_{S_p} (\br - \bq_p) \times \blambda(\br) \,\dd S_{\br} = \btau_p,
}
where the hydrodynamic traction, $-\blambda$, exerted on the colloids by the flow balances 
the non-hydrodynamic forces and torques, $\bbf_p$ and $\btau_p$, acting on the colloids  \cite{Pozrikidis1992}.

The Eqs.\ \eqref{eq:Stokes}-\eqref{eq:balanceT_continuum} form a linear system that can be solved for the fluid and colloid velocities.
The background flow is easily applied by splitting the flow velocity into a background and perturbation term $\bv = \bv_0 + \bv^{\star}$.
As the background velocity obeys the Stokes equation the perturbation term can be solved using Eqs.\ \eqref{eq:Stokes}-\eqref{eq:balanceT_continuum}
for $\bv^{\star}$ with the known \emph{slip} term $-\bv_0$ on the right hand side of \eqref{eq:no-slip-continuum} and the condition $\bv^{\star}(\bx) = 0$ at infinity.
In principle, any numerical method to solve the Stokes equation could be used to simulate a suspension of star colloids \cite{Corona2017}.
However, standard boundary integral methods are quite expensive for slender bodies as they require many quadrature points per body \cite{Koens2018}.
Instead, we use the rigid multiblob method, see a summary in \ref{sec:spheres} \cite{Usabiaga2016, Sprinkle2017}. 
The colloids are discretized with a small number of spherical \emph{blobs} along the arms as shown in Fig.\ \ref{fig:models}.
These blobs interact hydrodynamically through a regularized Green's function of the Stokes equation, the Rotne-Prager mobility tensor \cite{Rotne1969, Wajnryb2013}.
With this approach the long range hydrodynamics interaction are accurately captured
while near hydrodynamics are only captured with a coarse-grained model that neglects lubrication \cite{Usabiaga2016}.
Nevertheless the method is able to reproduce the viscosity of spherical colloid suspension for a properly chosen resolution, see Fig.\ \ref{fig:eta_sphere}. 
We compute the hydrodynamic interactions with a Fast Multipole Method which has a linear computational cost in the number of blobs in the system \cite{Yan2018a, Yan2020}.
With this approach we can simulate suspensions with up to $5000$ star colloids in a single CPU, see Fig.\ \ref{fig:models} bottom (Multimedia available online).

We place $N$ star colloids of the same kind in a simulation box of dimensions $L_x=L_y=5\si{\mu m}$ and $L_z=4.5\si{\mu m}$.
We solve the Stokes equations with semi-periodic boundary conditions,
i.e.\ we apply periodic boundary conditions in the $x$ and $y$ directions while the flow is unbounded in the $z$ direction.
We choose this setup because it is compatible with the background linear shear flow $\bv_0(\bx) = \gammaDot_0 z \what{\bs{e}}_x$.
To apply the semi-periodic boundary conditions we relay on the Fast Multipole Method developed by Yan and Shelley \cite{Yan2018a},
while to keep the colloids from diffusing to infinity along $z$ we include a short repulsive potential with the planes $z=-L_z/2$ and $z=L_z/2$.
We also introduce a purely repulsive potential between blobs.
Both potentials have the same functional form 
\eqn{
  \label{eq:potential}
  U(r) = \left\{\begin{array}{lc}
  \fr{U_0\delta}{\kappa}\pare{1 + \fr{\kappa}{\delta} + \fr{\delta}{2\kappa}} - \fr{U_0 r}{\kappa} \pare{1 + \fr{\delta}{\kappa} -  \fr{r}{2 \kappa} }, & r < \delta, \\
  U_0 \exp\pare{- \fr{r - \delta}{\kappa}}, & r \ge \delta.
  \end{array}\right.
}
The parameter $\delta$ is the distance when blobs start to overlap, we use $\delta=2a$ for blob-blob interactions and $\delta=a$ for blob bounding planes interactions, where $a$ is the blob radius.
The parameters $U_0=2\kt$ and $\kappa=0.1a$ control the magnitude of the force and the decay length.
These parameters only allow small overlaps between blobs while the interactions decay fast with the distance.
Note that since our model does not include lubrication, it is crucial to include a repulsive potential between colloids.

\begin{figure}
  \begin{center}
    \includegraphics[width=0.8 \columnwidth]{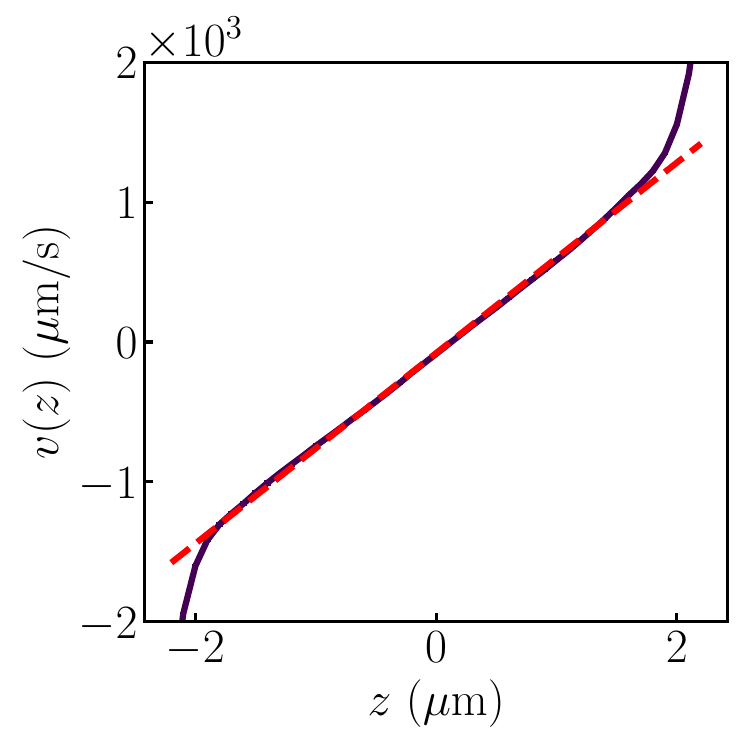}
    \includegraphics[width=0.8 \columnwidth]{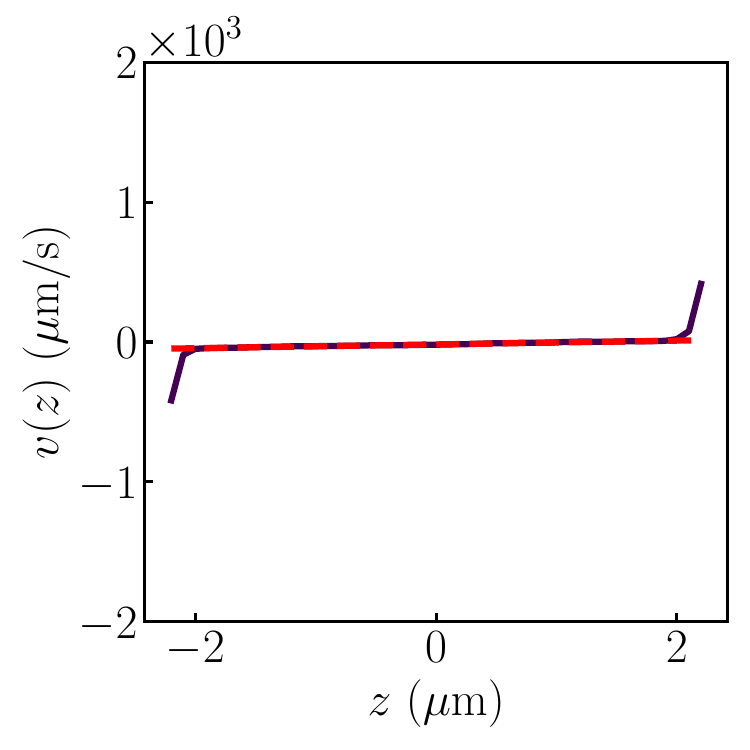}
    \caption{Average velocity profiles (continuous lines) and fits (dashed lines) for high Péclet and
      volume fractions $\phi=0.15 \text{ and }0.25$ (from top to bottom) for the model 3 of the star colloids.
      Away from the interfaces the velocity profiles are linear.     
    }
    \label{fig:profiles}
  \end{center}
\end{figure}

\section{Results}
\label{sec:results}
As explained in the previous section the suspension is probed by the background shear flow $\bv_0(\bx) = \gammaDot_0 z \what{\bs{e}}_x$,
with imposed shear rate $\gammaDot_0$. 
As the colloids increase the hydrodynamic drag the velocity profile within the suspension is modified. 
However, it remains linear with a slope $\gammaDot < \gammaDot_0$ as shown in Figs.\ \ref{fig:profiles} and \ref{fig:models} bottom (Multimedia available online). 
This slope can be measured and used to extract the suspension viscosity. 
In particular, the jump in stress across the interface relates the shear rate within the suspension with its viscosity \cite{Landau1987,Young2021}. 
Assuming that the surface tension is negligible, as for colloidal suspensions \cite{Aarts2004a,Aarts2004}, the jump of stress across the interface is 
\eqn{
  \label{eq:jump_stress}
  \llbracket \sigma_{xz} \rrbracket = \eta \fr{\partial v_x}{\partial z}\bigg\vert_{\text{interface}^-}  -
      \eta_0 \fr{\partial v_{x}}{\partial z}\bigg\vert_{\text{interface}^+} = 0.
}
From \eqref{eq:jump_stress} we extract the viscosity of the suspension as
\eqn{
  \label{eq:viscosity}
  \eta = \fr{\gammaDot_0}{\gammaDot} \eta_0.
}
We use this formula through the paper to measure the suspension viscosity
for the four colloid models at different volume fractions ($\phi$) and Péclet (Pe) numbers,
defined here as $\text{Pe}=\gammaDot d^2 / D_0$, where $D_0$ is the colloidal diffusion coefficient at $\phi=0$ and $d$ the colloidal diameter. 
The value of $\phi$ is computed with the solid volume fraction occupied by the colloids.
The volume of a colloid is the volume of its blobs subtracting the small overlaps of the blobs forming the arms.

\subsection{Viscosity versus volume fraction}
\label{sec:eta_vs_phi}
First we study how the concentration and type of colloid affects the suspension viscosity at high Péclet numbers.
We fix the temperature at $T=300K$, the solvent viscosity at $\eta_0 = 10^{-3} \si{mg / (\mu m \cdot s)}$,
like water, the shear rate at $\gammaDot_0=10^4\,\si{s^{-1}}$ and the system size $L_x=L_y=5\,\si{\mu m}$ and $L_z=4.5\,\si{\mu m}$.
With these parameters the Péclet number within the suspension is about $\text{Pe} \approx 100$.
We vary the number of colloids in the computational domain to measure the viscosity from dilute suspensions, with volume fraction $\phi=0.01$,
up to moderate concentrations at $\phi=0.30$.
At the highest volume fraction our simulations have $5229$ colloids for the star colloid without hooks and $2065$ for the largest colloid model.

\begin{figure}
  \begin{center}
    \includegraphics[width=0.85 \columnwidth]{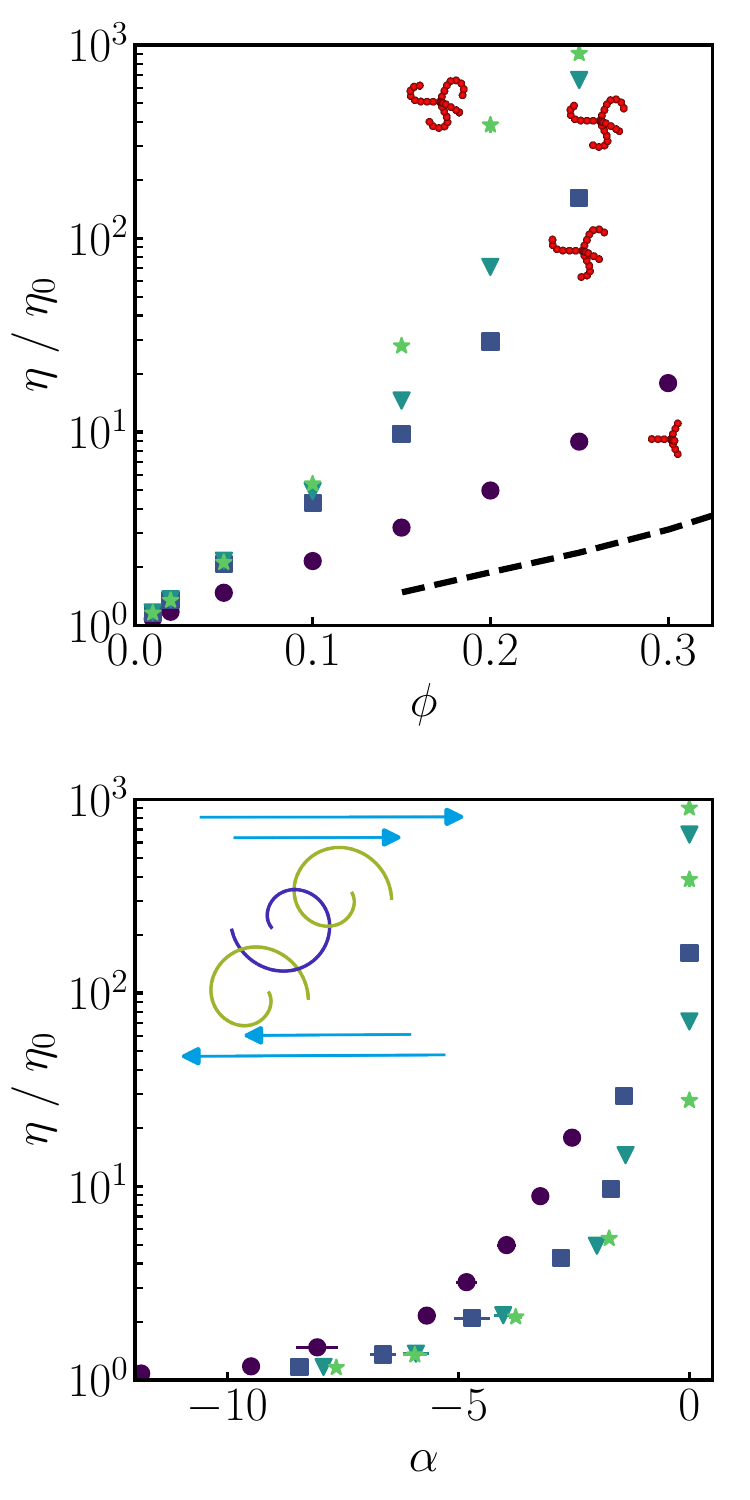}    
    \caption{{\bf (Top)} Viscosity increase ratio $\eta/\eta_0$ with volume fraction $\phi$ for four models of star colloids at high Péclet number ($\text{Pe}\sim 100$).  
      Colloids with curved arms or hooks show a dramatic viscosity increment at moderate volume fractions.  
      For comparison, we show the viscosity of a suspension of spherical colloids computed by Sierou \& Brady as a dashed line \cite{Sierou2001}.  
      {\bf (Bottom)} Viscosity versus exponent $\alpha$ obtained from the fit of the cluster size to a power law $P(N)\sim N^\alpha$.   
      The inset shows a sketch of a colloidal cluster in a shear flow.  
  }
  \label{fig:eta_vs_phi}
  \end{center}
\end{figure}

\begin{figure*}
  \begin{center}
    \includegraphics[width=0.4 \textwidth]{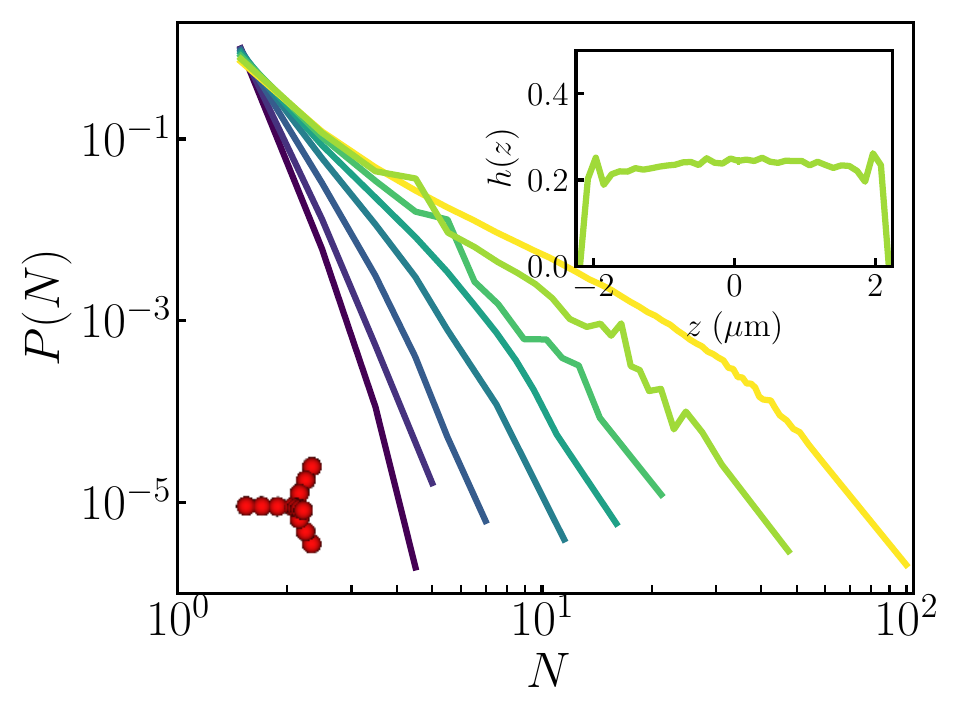}
    \includegraphics[width=0.4 \textwidth]{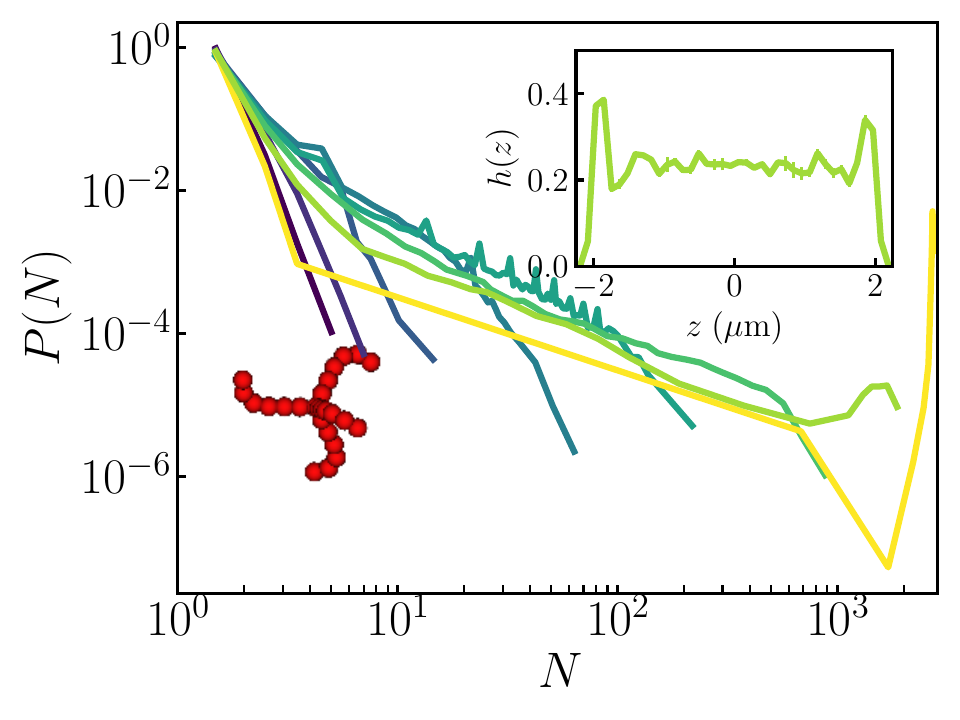}
    \includegraphics[width=0.4 \textwidth]{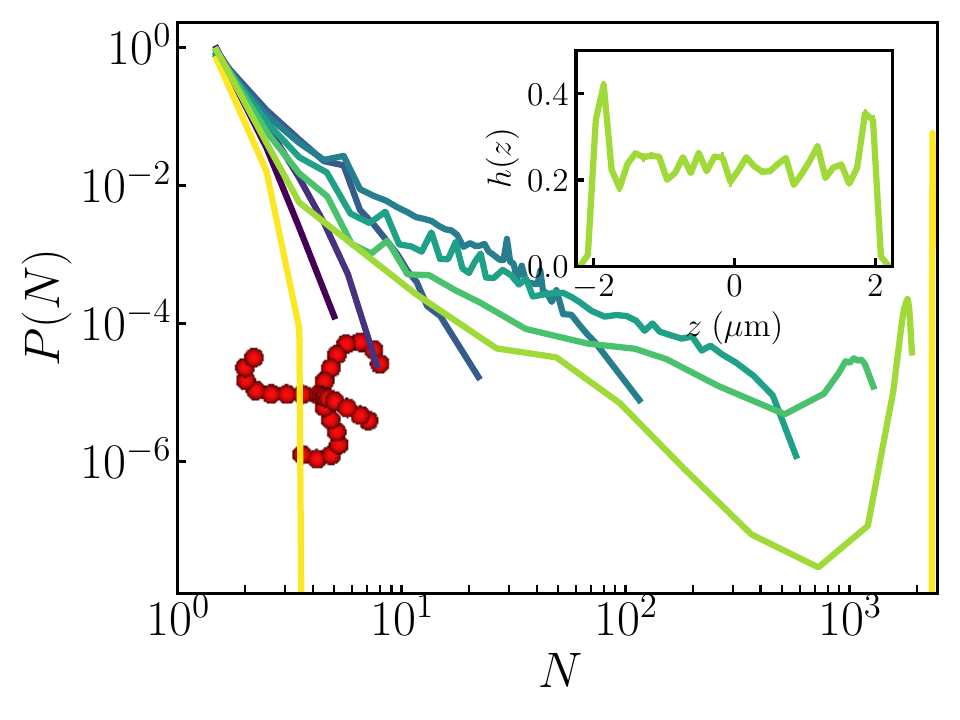}
    \includegraphics[width=0.4 \textwidth]{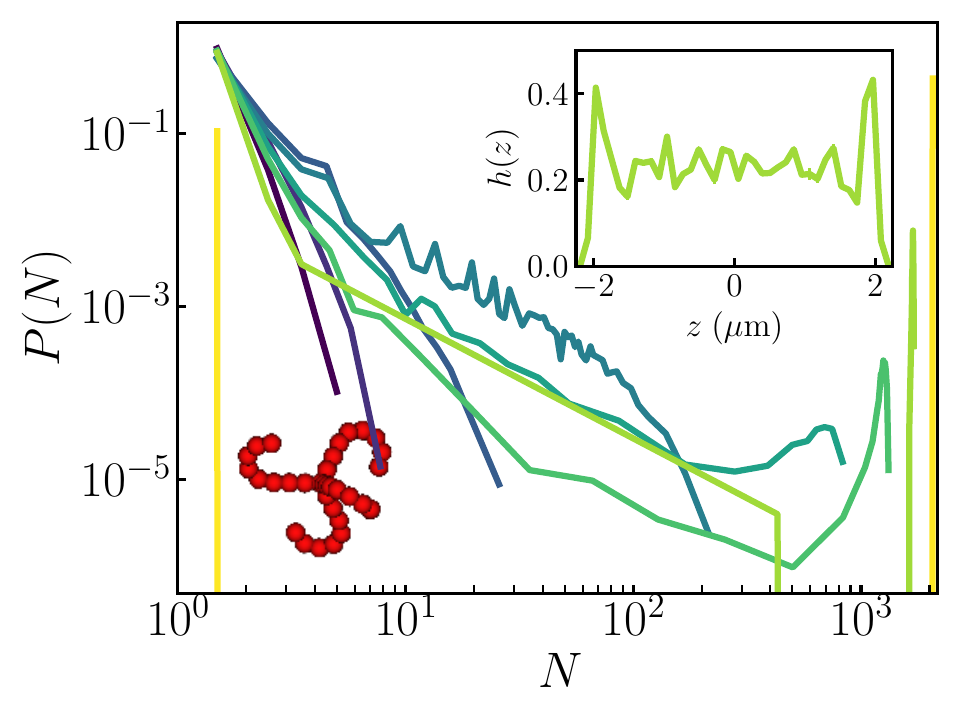}
    \caption{Histograms of cluster size at high Péclet number for the four colloid models
      at volume fractions $\phi=\{0.01,\, 0.02,\, 0.05,\, 0.1,\, 0.15,\, 0.20,\, 0.25,\, 0.30\}$.
      The curve colors indicate different volume fractions raising from dilute, for dark curves, to moderate concentrations for light curves.
      The histogram decays like a power law, $P(N)\sim N^{\alpha}$, for low concentrations.
      When the colloids have hooks the system percolates at moderate concentrations and a single cluster span the whole system.
      The critical volume fraction for percolation are  $\phi=0.25,\, 0.2,\, \text{ and } 0.15$ for the colloids with small, medium-sized, and large hooks respectively.     
      Colloids without hooks, top left, do not percolate at any concentration.     
      The insets show the concentration along the $z$-axis for suspensions at $\phi=0.25$.
    }
    \label{fig:clusters_phi}
  \end{center}
\end{figure*}

The viscosity values  versus volume fraction are shown in Fig.\ \ref{fig:eta_vs_phi} top. 
For the first model (no hooks) the viscosity increases faster than for spherical colloids but it only reaches moderate values, $\eta / \eta_0 \approx 19$ at $\phi=0.30$. 
However, for the other models the viscosity increase is dramatic. 
The viscosity raises more than a hundred times for the second model and near a thousand times for the third and fourth models at the same volume fraction. 
In contrast, in a suspension of spherical colloids the relative viscosity only increases about a factor two at similar volume fractions, see dashed line in Fig.\ \ref{fig:eta_vs_phi} top.

\begin{figure}
  \begin{center}
    \includegraphics[width=0.82 \columnwidth]{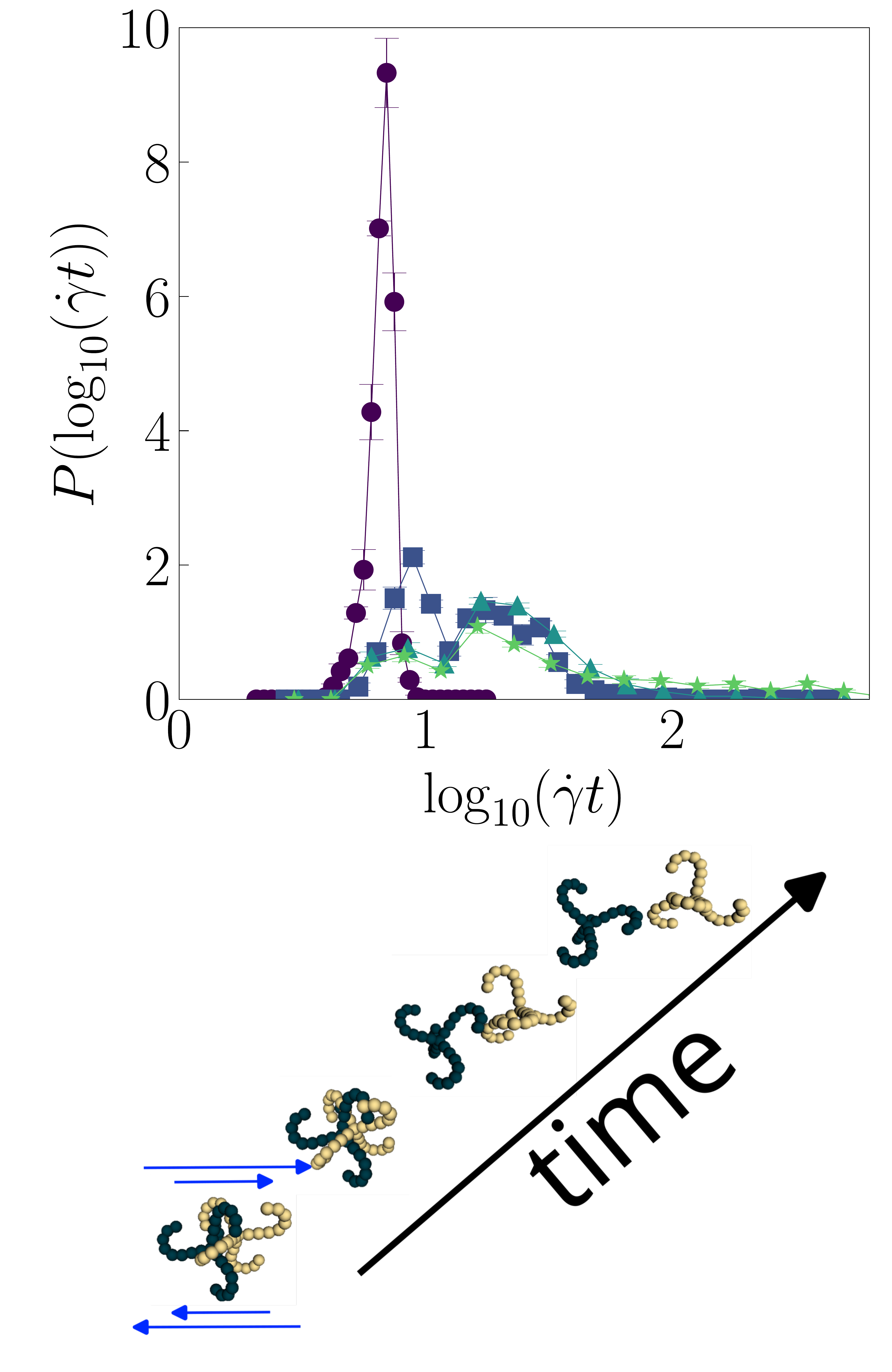}
    \caption{ 
      {\bf (Top)} Escape times of a two colloid cluster initially entangled with random orientations for the four colloid models
      represented with circles, squares, triangles  and stars respectively.
      The results are computed for $\text{Pe}=\infty$, i.e.\ without thermal fluctuations using $1000$ independent simulations.
      The presence of hooks modify completely the escape times at high Pe.
      {\bf (Bottom)} Snapshots of a cluster rotating in a shear flow (blue arrows) and eventually breaking apart (Multimedia view).
    }
    \label{fig:escape_times}
  \end{center}
\end{figure}

Dramatic viscosities increases as these has been observed experimentally in systems with elongated particles and friction contacts \cite{Bourrianne2022, Oseli2021}.
However, we observe similar increases without including any friction model in the simulations.
To understand the rheology we analysize the microstructure in the suspensions.
First, we compute the colloidal concentration along the $z$-axis, see insets in Fig.\ \ref{fig:clusters_phi}.
The confinement potential at the two planes $z=\pm L_z/2$ introduces a layering effect near them.
A phenomenon observed for other suspensions near hard walls \cite{Yeo2010a,Fernandez2012,Bian2014a, Vazquez-Quesada2023}.
However, away from the interface the colloidal concentration is constant and the suspensions are approximately homogeneous.
Then, we analyze the microstructure within the suspension through the presence of colloidal clusters.
We define a cluster as a set of colloids that are in contact, i.e.\ two colloids belong to the same cluster when the smallest distance between their blobs is $r_c \le 2a$.
Note that the soft repulsive potential between blobs allow small overlaps.
The histograms for cluster size are shown in Fig.\ \ref{fig:clusters_phi}.
For colloids without hooks (first model) the number of clusters decays with size like a power law, $P(N) \sim N^\alpha$, at all concentrations.
Interestingly, this is not the case for colloids with hooks.
In those cases the number of cluster decays like a power law a small concentrations.
However, once a critical volume fraction is reached, the system starts to percolate and a box-spanning cluster is formed.
For the second colloid model this starts to happen around $\phi=0.25$ while for the third and fourth models it occurs at $\phi=0.20$ and
$\phi=0.15$ respectively.
The formation of percolating clusters is a many body phenomena.
However, the underlying mechanism for the cluster formation can be explained with a two particle system.
We simulate two colloids initially entangled in a shear flow and we measure the time to disentangle. 
In Fig.\ \ref{fig:escape_times} (Multimedia available online) we show that while colloids with no hooks move apart immediately, colloids with hooks can remain entangled for very long times.
This phenomenons repeated over the whole suspension in many body simulations explain the different cluster size distribution for different cluster types.

To relate the presence of cluster to the suspenion viscosity we fit the cluster size histograms to a power law to extract the exponent $\alpha$.
When a system percolates the histogram is not well fitted by a power law, thus, we simple set $\alpha=0$ to indicate that the histogram does not decay with the cluster size.
We show, for the four colloid models, the viscosity versus the exponent $\alpha$ in the Fig.\ \ref{fig:eta_vs_phi} bottom.
We observe an approximate collapse of the curves which indicates the important role of the cluster size on the suspension viscosity.
The exponent characterizes the connectivity of the clusters within the suspension.
When the colloids possess hooks they can form large clusters able to sustain large stresses which increases the viscosity.
This result is similar to the large viscosity observed in suspensions where frictional contacts are active \cite{Seto2013, Mari2015, Bourrianne2022, Oseli2021}.
In those systems the frictional, non-hydrodynamic, interactions between particles generate a network that can resists the flow very efficiently.
Here we observe a similar effect generated by the colloids shape even with frictionless contacts.
It seems that similar to particle models with Coulomb's friction \cite{Seto2013, Ruiz-Lopez2023},
irregular shapes inhibit relative motion providing  additional constraints and frustrating the system.
This is also similar to the model of Jamali \& Brady  where \emph{explicit roughness}
of frictionless spheres was shown to lead to enhanced viscosities \cite{Jamali2019}.

The physical mechanism is sketched in the inset in Fig.\ \ref{fig:eta_vs_phi} bottom.
The shear flow tries to rotate the central colloid but the links to neighbor clusters prevent its rotation.
If the cluster is formed by just three colloids they could be rotated by the flow.
However, if they are part of a cluster that percolate the system they will not be able to rotate and will sustain a high stress which we measure as an increase in the viscosity.

\subsection{Viscosity versus Péclet number}
\label{sec:eta_vs_Pe}

\begin{figure}
  \begin{center}
    \includegraphics[width=0.9 \columnwidth]{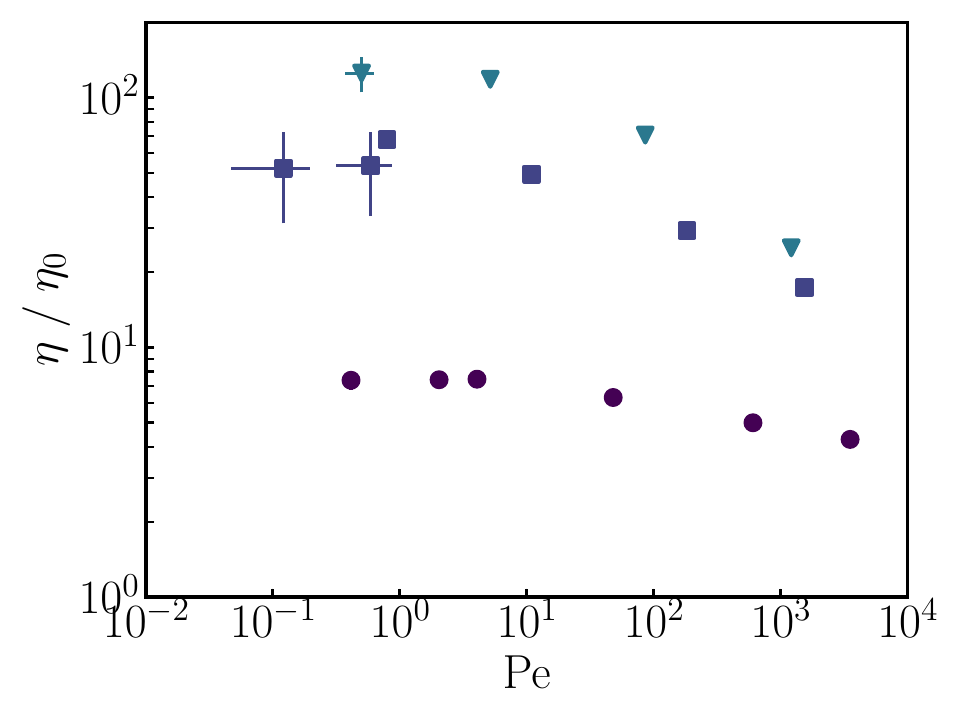}
    \caption{Viscosity versus Péclet number for three models of star colloids.
      For high Pe the viscosity show shear thinning and a plateau at low Pe numbers.
    }
    \label{fig:eta_vs_Pe}
  \end{center}
\end{figure}

Once established the viscosity growth with the volume fraction we focus on the viscosity dependence with the shear rate.
These simulations can be very expensive as there can be a large separation of time scales between the steric and the shear time.
The steric time, $\tau_{\tex{steric}} = \kappa / v = \kappa^2 / (\mu U_0)$, depends on the steric potential characteristic length, $\kappa$,
and a typical velocity that can be estimated as $v=\mu U_0 / \kappa$ where $\mu$ is the mobility of a single colloid and $U_0 / \kappa$ the typical force between colloids.
This time scale should be short enough to avoid the crossing between blobs at large shear rates,
i.e.\ $\tau_{\tex{steric}} < \tau_{\gammaDot} = 1 / \gammaDot$ for all shear rates.
Thus, the time scale separation grows as the imposed shear rate is reduced.
In order to reduce the computational cost of the simulations we exclude from this set of simulations the colloidal model with the largest hooks
and, we focus on the other three models.
We fix the volume fraction at $\phi=0.2$ and vary the imposed background shear rate, $\gammaDot_0$, while keeping the other parameters constant.
We explore Péclet numbers in the range $\text{Pe} \approx \pare{10^{-1}\; \dash\; 5 \cdot 10^{3}}$.

The viscosity values are shown in Fig.\ \ref{fig:eta_vs_Pe}.
At high Pe the viscosity shows a clear shear thinning for all the models considered while at low shear rates the viscosity reaches a plateau.
The shear thinning is explained by a combination of factors.
First, we include a soft repulsive potential between colloids which
has been shown to introduce a shear thinning on suspensions of spherical colloids \cite{Maranzano2001, Nazockdast2012, Swan2013}.
As the colloids repel each other the effective free space for them is lower.
However, this excluded volume depends on the shear pushing the colloids against each other.
Higher shear rates can overcome the steric repulsion which increases the volume accessible to the colloids and thus reduces the
\emph{effective} volume fraction occupied by the colloids and thus the viscosity \cite{Mari2014, Chatte2018}.
Second, the Brownian motion contribution to the viscosity decreases with the Pe number,
since when the time scales associated with the flow are shorter than the Brownian ones, the thermal motions cannot generate deformations that contribute to the
stress and thus the viscosity \cite{Bossis1989, Brady1997, Foss2000}.
Finally, two mechanisms that counter shear thinning, leading to shear thickening, are the lubrication interaction between colloids \cite{Foss2000}
and frictional interactions \cite{Seto2013, Mari2014}.
However, as we explained in Sec.\ \ref{sec:setup} our numerical method does not include lubrication
nor explicit frictional interactions and thus these competing mechanism to the shear thinning are not present in our numerical results.
Interestingly, the shear thinning is more pronounced for colloids with hooks than for the simpler straight star colloids as shown in Fig.\  \ref{fig:eta_vs_Pe}.
Similar shear thinning results were found for suspensions of Czech hedgehog colloids, star colloids without hooks,
by Westwood et al., who used a close related numerical method \cite{Westwood2022}.
However, here the presence of the hooks enhance the suspension viscosity dramatically over a wide range of Pe numbers.

In the work of Bourrianne et al.\ also strong thinning was reported in non-frictional (hydrophobic) fumed silica particles,
where the presence of a yield stress was discovered \cite{Bourrianne2022}.
The later phenomenon was hypothesised to be connected to the presence of a percolating cluster at low Pe which is compatible with the present study.
This could be another contributing factor to the thinning reported here.
However, to clearly verify its nature a small amplitude oscillatory (SAOS) study should be done to check elastic versus viscous contributions to the stress
which will be done in a future work.

\section{Conclusions}
\label{sec:conclusions}

We have established through numerical simulations that suspensions of star colloids have extremely large viscosities even at moderate volume fractions.
The relative viscosity can increase up to a factor 1000 at a volume fraction of $\phi=0.25$ in these suspensions
while with spherical colloids the relative viscosity only doubles at similar volume fractions.
A cluster analysis reveals that the viscosity increases when colloids form large clusters that can resist stresses.
The results in Fig.\ \ref{fig:eta_vs_phi} show that the viscosity is correlated with the cluster size through the exponent $\alpha$
of the probability distribution function, $P(N) \sim N^{\alpha}$, of the cluster sizes.
The largest viscosities are observed for $\alpha \approx 0$, when most colloids form a single cluster that spans the whole domain.

The formation of clusters is controlled by the volume fraction and, importantly, the colloid shape. 
Colloids with hooks can entangle with their neighbors which favors the formation of large clusters. 
Theoretical and experimental works have shown that the viscosity in colloidal suspensions can be enhanced by introducing friction contacts between colloids \cite{Seto2013, Mari2015, Tanner2016, Bourrianne2022}.
Here we show that a similar enhancement  can by controlled by the colloid shape instead of by their surface chemistry interactions.

In particular, we have shown that the rheology of colloidal suspensions can be affected by small changes in the colloid shape.
Hooked colloids lead to much larger viscosity increases and, additionally, to a stronger shear thinning.
This is relevant now that complex-shaped colloids can be designed and synthesized with an array of methods.
Combining the colloidal geometry with their surface chemistry would allow to design colloids that show different shear thinning or thickening effects at different shear regimes.
Finally, star colloids can be chiral, i.e.\ they may not be symmetric under reflections.
The chirality of the colloids could affect their rheology as flowing in some directions or under some  drives could be easier than under others.
It could be explored if this is the case and how it can be exploited to explore new physics in passive suspensions \cite{Zhao2022}.

\section*{Acknowledgments}
\label{sec:acknowledgments}
The project that gave rise to these results received the support of a fellowship from ``la Caixa''
Foundation (ID 100010434), fellowship LCF/BQ/PI20/11760014, and from the European Union's Horizon 2020 research and innovation
programme under the Marie Skłodowska-Curie grant agreement No 847648. 
Funding provided by the Basque Government through the BERC 2022-2025 program
and by the Ministry of Science and Innovation: BCAM Severo Ochoa accreditation
CEX2021-001142-S/MICIN/AEI/10.13039/501100011033 and the project PID2020-117080RB-C55
``Microscopic foundations of soft matter experiments: computational nano-hydrodynamics (Compu-Nano-Hydro)'' are also acknowledged.

\section*{Data availability}
The data that support the findings of this study are available
from the corresponding author upon reasonable request.

\appendix

\section{The rigid multiblob method}
\label{sec:spheres}
\begin{figure}
  \begin{center}
    \includegraphics[width=0.85 \columnwidth]{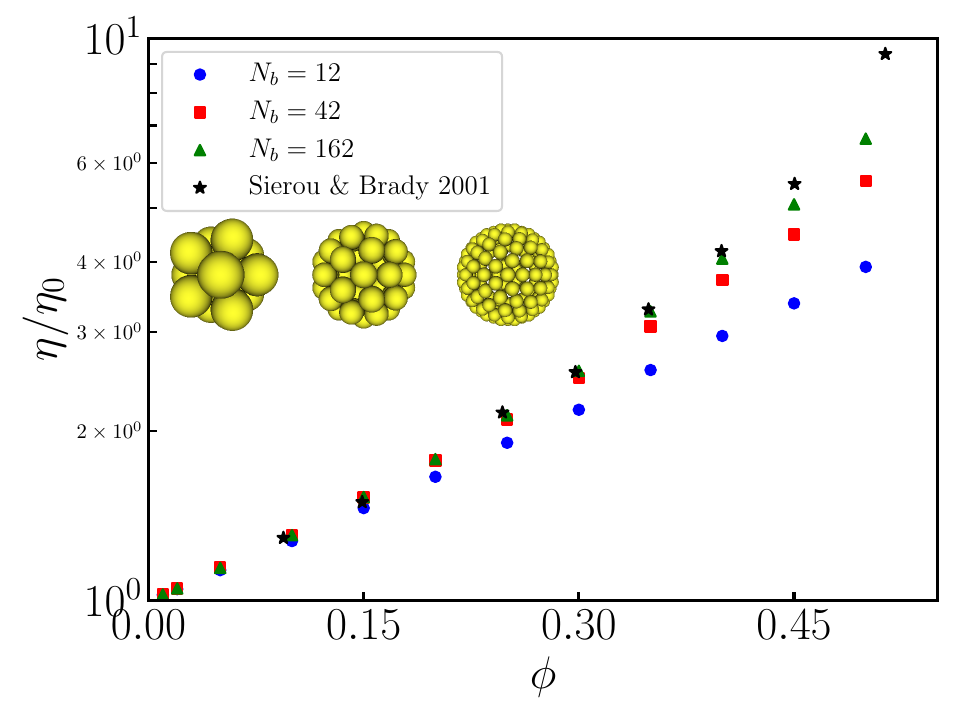}
  \caption{
    Viscosity of a suspension of spherical colloids versus volume fraction.
    We discretize the colloids with $N_b$ number of blobs, see insets, and compare with the results of Stokesian dynamics from Sierou \& Brady \cite{Sierou2001}.
    The viscosity values are under predicted due to the lack of lubrication in our numerical method.
    The discrepancy decreases for higher resolutions.
  }
  \label{fig:eta_sphere}
  \end{center}
\end{figure}

We delineate here the rigid multiblob method while a detailed description can be found in Ref. \cite{Usabiaga2016}.
Colloids are discretized by a finite number of markers, or blobs, on its surface, see Figs. \ref{fig:models} and \ref{fig:eta_sphere}.
Then, the integrals on the balance of force and torque, Eqs.\ \eqref{eq:balanceF_continuum}-\eqref{eq:balanceT_continuum}, become sums over the blobs
\eqn{
\label{eq:balanceF}
\sum_{i\in \text{colloid } p} \blambda_i &= \bbf_p, \\
\label{eq:balanceT}
\sum_{i \in \text{colloid } p} (\br_i - \bq_p) \times \blambda_i   &= \btau_p.
}
Here $\blambda_i$ is the force acting on the blob $i$ to enforce the rigid motion of the colloid.
The \emph{slip} condition is evaluated at every blob $i$ in colloid $p$ as 
\eqn{
\label{eq:no-slip}
\bv(\br_i) =  \sum_j \bM_{ij} \blambda_j &= \bu_p + \bomega_p \times (\br_i - \bq_p) + \bu_s(\br_i) 
}
where the mobility $\bM$ couples the blobs hydrodynamically and we have included an slip term, $\bu_s(\br_i)$, that we will discuss in a moment.
The submatrix $\bM_{ij}$ computes the flow at the blob $i$ generated by the force acting on the blob $j$.
In boundary integral methods the mobility matrix would be the Green's function of the Stokes equation, $\bG(\bx, \by)$.
The rigid multiblob method uses instead a regularization of the Green's function, the so-called Rotne-Prager (RPY) mobility \cite{Rotne1969,Wajnryb2013}
\eqn{
  \label{eq:RPY}
  \bM_{ij} = \bM(\br_i, \br_j)  &=  \fr{1}{(4\pi a^2)^2}  \int  \delta(|\br' - \br_i|-a) \bG(\br', \br'') \nonumber \\
  & \delta(|\br'' - \br_j|-a) \dd^3 r'' \dd^3 r', 
}
which is written here as the double integral over the blobs surface.
The advantage of this approach is that the RPY mobility is positive definite, which makes the scheme robust even with very low resolutions and
eases the generation of the Brownian noise \cite{Westwood2022,Sprinkle2019,Delong2015b}. 
The equations \eqref{eq:balanceF}-\eqref{eq:no-slip} form a linear system for the colloidal linear and angular velocities and the blobs constraint forces.
We can write the linear system explicitly as
\eqn{
\label{eq:linear_system}
\left[\begin{array}{cc}
\bM & -\bK \\
-\bK^T & \bzero 
\end{array} \right]
\left[\begin{array}{c}
\blambda \\
\bU 
\end{array} \right] =
\left[\begin{array}{c}
\bu_s \\
-\bF 
\end{array} \right],
}
with the notation $\bU = \{\bu_p, \bomega_p\}$, $\bF=\{\bbf_p, \btau_p\}$ and $\blambda=\{\blambda_i\}$.
The geometrix matrix $\bK$ transform rigid body velocities to blobs velocities, thus we have used it to write the no-slip condition in \eqref{eq:linear_system},
i.e.\ $\pare{\bK \bU}_i = \bu_p + \bomega_p \times (\br_i - \bq_p)$ for blob $i$.
The slip term in the right hand side, $\bu_s = -\bv_0(\br) + \sqrt{2\kt / \dt} \bM^{1/2} \bW^n$, includes the imposed background flow and the Brownian noise.
In the second term $\bW^n$ is a vector of Gaussian noise generated at the time step $n$ and $\bM^{1/2}$ is the \emph{square root} of the mobility matrix (e.g.\ a Cholesky factorization).
This stochastic slip is equivalent to include the noise directly in the Stokes equation as in \eqref{eq:Stokes} \cite{Westwood2022,Sprinkle2019}.
With the appropriate stochastic integrator the solution of this linear equation reproduces the correct Brownian motion of the colloids \cite{Westwood2022,Sprinkle2019}.
We use the stochastic trapezoidal method described in Ref. \cite{Sprinkle2017}.

To validate our numerical method we compute the viscosity of a random suspension of spherical colloids for different volume fractions.
We use different resolution, i.e.\ number of blobs, to discretize the colloids and show the results in Fig.\ \ref{fig:eta_sphere}.
For low volume fraction, $\phi\approx 0.15$, we get good agreement with the results of Sierou \& Brady \cite{Sierou2001} with low resolutions,
for example using 12 blobs per colloid.
As the concentration is increased, and the near hydrodynamic interactions gain importance, we need to use higher resolutions the recover the correct viscosity.
The discrepancy can be made arbitrarily small with higher resolutions at the expense, of course, of higher computational cost.
As shown in Fig.\ \ref{fig:eta_sphere} the viscosities computed with the rigid multiblob method are underpredicted,
which is a general feature of our approach due to the lack of lubrication interactions in the rigid multiblob method.





\bibliography{Biblio}

\end{document}